\documentclass[a4paper,brazil,12pt,nofootinbib,aps,pra]{revtex4}
\usepackage[utf8]{inputenc}
\usepackage{float}
\usepackage[T1]{fontenc}
\usepackage{amsmath}
\usepackage{slashed}

\usepackage{braket}
\usepackage{nicefrac}
\usepackage{bbm}   
\usepackage{mathrsfs}
\usepackage{multirow}
\usepackage{graphicx}
\usepackage{xcolor}
\usepackage{subfigure}

\usepackage{caption}
\DeclareCaptionFont{xipt}{\fontsize{11}{13}\mdseries}
\usepackage[font=xipt,labelfont=bf]{caption}
\usepackage{setspace}  
\usepackage{moreverb}

\usepackage{hyperref}
\hypersetup{
citebordercolor=0 .75 1,
linktocpage=true,
backref=true}

\begin{document}
\title{
Curving Trajectories of Light Rays in Refraction Index Gradients
}

\author{O. Polachini }
\email{opolachini26@amherst.edu}

\author{F. Marques }
\email{fabriciomarques@if.usp.br}

\affiliation{Amherst College Physics and Astronomy Department, Amherst, Massachusetts, 01002, USA}
\affiliation{University of S\~ao Paulo, Institute of Physics, 66318, 05315-970, S\~ao Paulo, SP, Brazil}

\date{\today}

\begin{abstract}{
    \textbf{Abstract:} 
This work seeks to present an investigation about the trajectories followed by rays 
of light passing through refractive index gradients that was entirely carried by high 
school students. 
Such trajectories are curved, 
therefore contradicting the common sense that light should always travel along 
straight lines. This fact causes the formation of distorted and striated images, 
similarly to a type of mirage known as \emph{Fata Morgana}.
Using a rectangular aquarium containing solutions with different gradients of sugar 
or water with a temperature gradient, we analyzed the conditions for image inversion 
to occur. Also, we were able to reconstruct a distorted image through our theoretical 
predictions.

Trajectories of light rays. 
Refractive index gradient. 
Concentration gradient. 
Temperature gradient. 
Low cost experiments. 
Active learning. 
Scientific Olympiads.  
  }\end{abstract}

\maketitle

\newpage

\section{Introduction}
\label{sec:introducao}

As most of us know, mirages are an optical phenomenon caused by the refraction of light. 
They can be observed in nature but also easily produced in a laboratory in a controlled manner when our goal is to study specific properties concerning them. 
It is possible to classify mirages in two basic types according to the position of the image formed 
with respect to the object: superior and inferior. 
A \emph{superior mirage} forms an inverted image above the object, while an \emph{inferior mirage} also forms an inverted image, but, instead, below the object~\cite{Young2}. 
The term \emph{``Fata Morgana''} defines a specific type of superior mirage in which multiple 
distorted and striated images are produced. Those images, can change rapidly and alternate, in a 
complex way, between straight and inverted~\cite{Young1}. 
As they are significantly deformed vertically, it is common for them to cause confusion that will, 
for instance, lead observers in a beach to see a boat as if it was levitating over the water. Even 
after multiple attempts by scientists such as Pernter and Exner~\cite{Exner}, there is still no 
consensus on the exact cause of the formation of Fata Morgana mirages. However, it is theorized by 
physicist Andrew T. Young that such an effect is caused by a thermal inversion so intense that it 
causes a curvature of light rays greater than the curvature of the Earth~\cite{Young2}.

It is known that a gradient of refractive indices causes an effect that is similar to a mirage, in 
which the final formation of images can resemble that of a Fata Morgana~\cite{IYPT}. Therefore, this 
work seeks to replicate such phenomenon by means of an image reconstruction performed by studying the 
path of light rays through two gradients of refraction indices: one of them resulting from a 
temperature gradient, and the other, from a gradient of concentration of sugar particles.

This investigation was completed as part of the authors' participation on the International Young 
Physicists' Tournament (IYPT), a competition that seeks to encourage high school students to solve 
open physics problems. Such problems consist of small paragraphs defining a specific situation 
or phenomenon, and then establishing some task that will not have a final or closed answer but 
will lead students to find creative and deep explanations for that situation. Ordinary high school 
physics will certainly not be enough to accomplish those tasks and, therefore, they will learn much 
more than what is usually taught in regular curricula. Those are typical characteristics of an 
active learning method. In particular, when students at this age range are 
taught about the fundamentals of geometrical optics, they will be simply told that 
``light travels along straight paths'' in a sentence that, to be fair enough,
often finishes at least saying that it will be so ``provided that light is 
travelling through a homogeneous and isotropic media''
(although a more careful thought about the meaning of concepts like homogeneity and 
isotropy are not carried on in general and, many times, those words are just thrown 
in the wind).
Also, when students are 
presented to the phenomenon of refraction, it will be done through the introduction 
of the idea of two homogeneous and isotropic media meeting one another in a plane 
interface and, possibly, at most, the deviation of the transmitted ray with respect 
to the incoming one's path will be justified invoking Fermat's principle without 
further elaboration. 
Sometimes, demonstrations of situations deviating from this 
ideal set of simplifying hypothesis are carried, of course, but, many times, 
due to a general assumption that students at those grades will not be able to 
deal with more complicated tasks, perhaps because they probably lack essential 
mathematical knowledge, 
further investigation on off-ideal situations are regarded as non-practical and 
not significant.
Another common justification to avoid carrying a project involving further 
investigation on more complex phenomena is that students will have a limited 
amount of time to study before applying to universities and that such type 
of knowledge will not be required by some specific type of standardized test 
they will be required to take. 

As an opposition to this common sense, this work was carried by high school students under a teacher's supervision, as a research project within the context of a physics tournament, the IYPT. Despite the fact that results in this type of competition have been considered as a criteria for acceptance by an increasing number of universities over the years, it is undeniable that involving themselves with a research project, being able to communicate their results and to go as further as possible and, later, even publishing it, will constitute a unique advantage to their personal development and, also, will certainly be a truly significant knowledge.

This work is organized as follows. In Section~\ref{sec:fundamentos}, we will present the theoretical 
model relevant to our work. In Sections~\ref{sec:experimental}~and~\ref{sec:analise}, we will 
describe the equipments used and our experimental procedures, following to an analysis of the 
collected data, including a procedure to reconstruction of images. Finally, in 
Section~\ref{sec:conclusao}, we will present our final considerations. A more detailed discussion 
about the influence of the wavelength will be found in Appendix~\ref{subsec:A}.

\section{Theoretical Foundations}
\label{sec:fundamentos}

\subsection{Formation of Mirages in Nature}

\begin{figure}[H]
    \centering
    \includegraphics[%
    width=0.50\columnwidth]{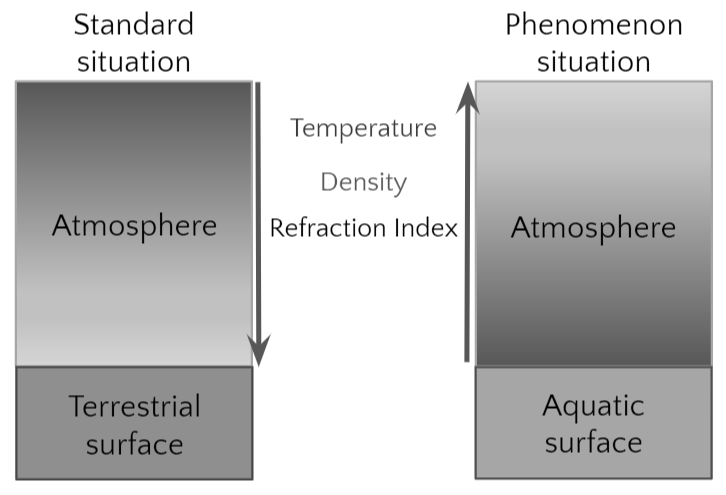}
    \caption{Comparison between the standard atmospheric configuration and the situation for which the Fata Morgana mirage occurs.}
    \label{fig:inversao}
\end{figure}

As mentioned in the previous section, the main cause for superior mirages is the inversion of air temperature.
In the most common situation, Earth will absorb solar radiation and transfer this energy in the form of heat to lower atmospheric layers. 
As air heats up, there is an increase in pressure and, consequently, an expansion. The direction of this expansion can be lateral or vertical; 
in vertical expansion, air that has warmed up near the surface will expand as it rises, while a layer of cooler and denser air just above descends to occupy the now less dense region left below. 
This is the general principle that governs a convection current. 
Of course, as higher altitudes are reached, air will become more rarefied and temperatures will drop again.
However, at lower altitudes, before air layers close to Earth can expand and exchange places with cold air above them, there will be a layer of lower and less dense warm air followed by layers of increasingly colder and denser air, forming a temperature and pressure gradient.
In a superior mirage we observe the opposite, as illustrated in Figure~\ref{fig:inversao}. 
This often occurs over water surfaces because water exchanges heat with air slower than land does.

Both density and temperature are relevant parameters to the formation of a gradient of refractive indices. 
Furthermore, as we demonstrate below, it is possible to consider that the relation between these quantities and the value of the refractive index is linear.

\begin{figure}[H]
    \centering
    \includegraphics[%
    width=0.60\columnwidth]{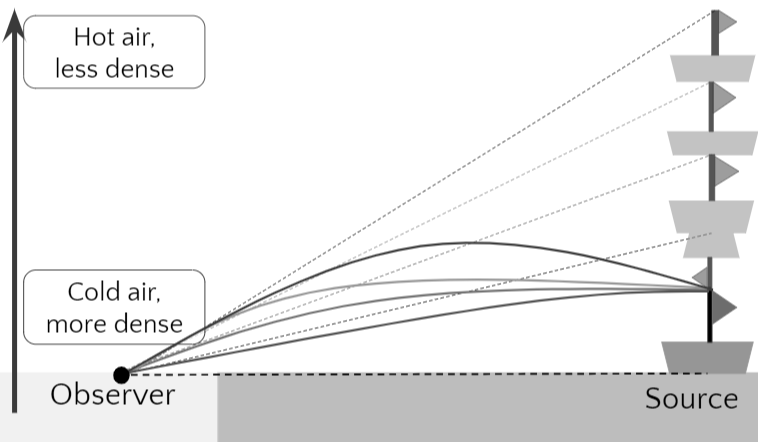}
    \caption{Graphic representation of curved paths followed by light causing a Fata Morgana mirage.}
    \label{fig:barco}
\end{figure}

Being easier for light to pass through less dense media -- for instance, hot air -- and, as there is a sequence of layers of different densities that form the gradient (whether due to differences of temperature or of particle concentration), an infinite number of refractions must to occur as light passes through different layers. 
When each layer is infinitesimal in width, that is, as we move to the continuum limit, the infinite sequence of refractions will be perceived as a curved trajectory followed by the light ray. 
A side effect of this phenomenon is that an observer can capture the light that emerged from the same point of a source, but has traveled through different paths, which creates the illusion of multiple replicas of that point. 
Thus, multiple images of the same object can be seen by this observer.
In Figure~\ref{fig:barco}, we illustrate this idea with the picture of a boat, which has its apparent size distorted and its image inverted in some of the illusions created by the curved paths followed by light rays.

\subsection{Trajectories of Light Rays}

\begin{figure}[H]
    \centering
    \includegraphics[%
    width=0.78\columnwidth]{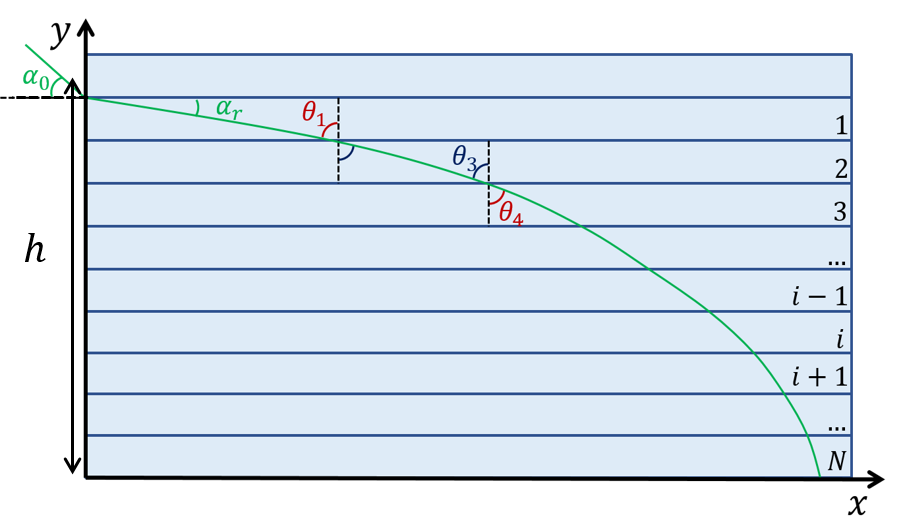}
    \caption{Graphic representation of the aquarium used in the experiments.}
    \label{fig:camadas}
\end{figure}

Figure~\ref{fig:camadas} shows a representation of the trajectory followed by a ray of light inside an aquarium.
The light passes through $N$ layers, each one having an infinitesimal thickness $\delta$ and a refraction index $n_i$ associated to it, where the index $i=1,2,...,N.$ labels the $i$-th layer. 

As we look more closely at the trajectory, we can write the boundary conditions associated with light bending as:


\begin{equation}
\left\{
    \begin{aligned}
        & y(x_0=0)=h     
        \\
        & \frac{dy}{dx}(x_0=0)  =  -\tan\alpha_r 
    \end{aligned}
    \right.
    \label{eq:contorno}
\end{equation}
where:

\begin{equation*}
    \sin{\alpha_r} = \frac{n_{\text{air}}}{n_1} \sin{\alpha_0}
\end{equation*}

The boundary conditions \eqref{eq:contorno} are obtained from the height $h$ where light incides, as measured with respect to the bottom of the aquarium, the angle $\alpha_0$ the light ray makes with the horizontal and may depend on a first refraction that occurs when the light passes through the aquarium's side wall. 
Such conditions are highlighted here, taking as parameters, for example, $h$ and $\alpha_0$.

Knowing that there is a linear relation connecting the refractive index of layers $n$ and their concentration $C$ and that $n$ is an increasing function of $C$, and also, that $C$ varies linearly and decreases with increasing heights $y$, measured from the bottom, we know that the relation between the refractive index $n$ of a layer and its height $y$ must be linear and decreasing, i.e.:

\begin{equation}
    n = n_0 - k y \; \Rightarrow \; \frac{dn}{dy} = -k
    \label{eq:indiceDeRefracaoAltura}
\end{equation}
where $k$ is a constant.

\begin{figure}[H]
    \centering
    \includegraphics[%
    width=0.38\columnwidth]{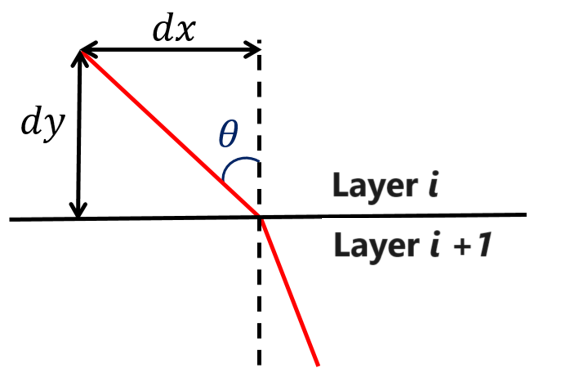}
    \caption{Geometry of the transition of a light ray from layer $i$ to $i+1$.}
    \label{fig:angulo}
\end{figure}

Figure~\ref{fig:angulo} gives, imediately:

\begin{equation}
    \frac{dy}{dx} = - \cot \theta
    \label{eq:dydx}
\end{equation}

From Snell's law, we have:

\begin{equation}
        n_1\sin{\theta_1} = n_2\sin{\theta_2} = ... = \text{const.} 
        \Rightarrow
        n(y) = \frac{\text{const.}}{\sin{\theta}}
        \label{eq:Snell}
\end{equation}

Differentiating the above relation with respect to $\theta$, we find:

\begin{equation}
        \frac{d}{d\theta} \left[ {n(y) = \frac{\text{const.}}{\sin{\theta}}} \right] 
        \Rightarrow
        \frac{dn}{d \theta} 
        = - \text{const.} \frac{\cot \theta}{\sin \theta}
    \label{eq:SnellDerivada}
\end{equation}
and, using \eqref{eq:indiceDeRefracaoAltura} and \eqref{eq:dydx} together, we can manipulate the derivative with respect to $\theta$ of Snell's law expression to get to a differential equation for the trajectory followed by a light ray. 

According to Figure~\ref{fig:angulo}, we have:

\begin{equation*}
    \sin \theta 
    = \frac{dx}{\sqrt{dx^2 + dy^2} }
    \Rightarrow 
    \frac{d}{dx} \left[ 
    \sin{\theta} 
    =  \frac{dx}{\sqrt{dx^2 + dy^2}} 
    \right] 
    \Rightarrow 
    \frac{1+{\frac{dy}{dx}}^2}{\frac{d^2 y}{dx^2}} 
    = \frac{dx}{d\theta}
\end{equation*}

Combining the above relation with
\eqref{eq:indiceDeRefracaoAltura}, \eqref{eq:dydx}, \eqref{eq:Snell} and \eqref{eq:SnellDerivada}, we can write:

\begin{equation*}
      \frac{1}{n} \frac{dn}{d \theta} 
      =\frac{1}{n} \frac{dn}{dy} \frac{dy}{dx}\frac{dx}{d\theta}
      = - \cot \theta
      \Rightarrow
       \left( \frac{1}{n} \right) \left( -k \right) \left(-\cot{\theta} \right) 
       \left( \frac{1+{\frac{dy}{dx}}^2}{\frac{d^2 y}{dx^2}} \right) 
       = -\cot{\theta}
\end{equation*}
and, after a few manipulations, we will get to the following differential equation:

\begin{equation}
    \begin{aligned}
        \frac{-k}{[n_0 - ky]}\left(1+ \left(\frac{dy}{dx} \right)^2 \right) = \frac{d^2 y}{dx^2}  \Rightarrow  y''(x) + \frac{-k}{[n_0 - ky(x)]}\left( 1+y'(x)^2 \right)=0
    \end{aligned}
    \label{eq:trajetoria_y}
\end{equation}

It will be simpler to work with this equation if we change variables from $y(x)$ to $n(x)$. By doing this, equation \eqref{eq:trajetoria_y} becomes:

\begin{equation}
    -\frac{n''(x)}{k} + \frac{k}{n} \left( 1+\frac{n'(x)^2}{k^2} \right) = 0 \Rightarrow n''(x)n(x) - n'(x)^2 = k^2
    \label{eq:trajetoria_n}
\end{equation}

It is easy to check that equation \eqref{eq:trajetoria_n} has a solution of the form $n(x) = Ae^{\rho x}+ Be^{-\rho x}$, with $\rho = \frac{k}{\sqrt{2AB}}$, and with $A$ and $B$ constants to be fixed by boundary conditions \eqref{eq:contorno}. However, we must first adapt those conditions to the new variables. This gives:

\begin{equation}
\left\{
    \begin{aligned}
        & y(x_0=0)=h     
        \\
        & \frac{dy}{dx}(x_0=0)  =  -\tan\alpha_r 
    \end{aligned}
    \right.
    \Rightarrow
\left\{
    \begin{aligned}
        & n(x_0=0) = n_0 - kh = A+B \\
        & n'(x_0=0) = k \tan{\alpha_r} = \rho(A-B)
    \end{aligned}
    \right.
\end{equation}
and, solving for $A$ and $B$, we find:

\begin{equation*}
    A = \frac{1}{2} \left( n_0 - kh \right) (1+\sin{\alpha_r})
    \quad \text{and} \quad
    B = \frac{1}{2} \left( n_0 - kh \right) (1-\sin{\alpha_r})
\end{equation*}
with:

\begin{equation*}
     \rho = \frac{k}{\cos{\alpha_r}(n_0 -  kh)}
\end{equation*}

In this way, the analytical solution of equation \eqref{eq:trajetoria_n} is found to be:

\begin{equation}
    n(x) = \frac{\left( n_0 - kh \right)}{2} \bigg[ (1+\sin{\alpha_r}) e^{\rho x} + (1-\sin{\alpha_r}) e^{\rho x} \bigg]
\end{equation}

Finally, we just have go back to the variable $y(x)$ – which is the height of a point of the trajectory followed the light ray as a function of the horizontal distance travelled by light. The result depends on the first refracted angle $\alpha_r$, the index of refraction of the first layer of incidence $n_0$, the rate of change of the index of refraction with height $k$ and the height of incidence $h$, as we see:

\begin{equation}
    y(x) = \frac{n_0}{k} - \frac{\left( n_0 - kh \right)}{2k} \bigg[(1+\sin{\alpha_r})e^{\rho x} + (1-\sin{\alpha_r})e^{-\rho x} \bigg]
\end{equation}

\section{Experimental Procedure}
\label{sec:experimental}

\begin{figure}[H]
	\begin{center}
	\subfigure{\includegraphics[width=0.7\linewidth]{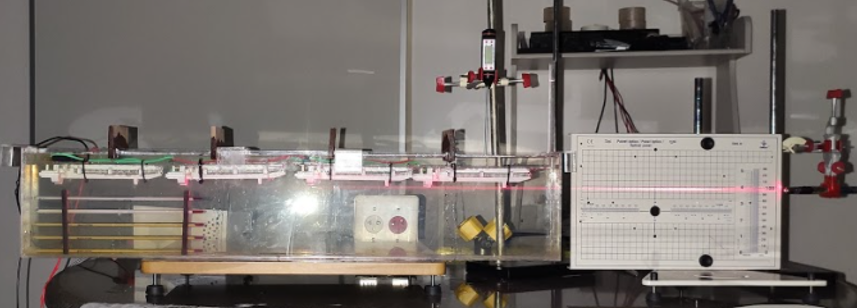}}
	\subfigure{\includegraphics[width=0.7\linewidth]{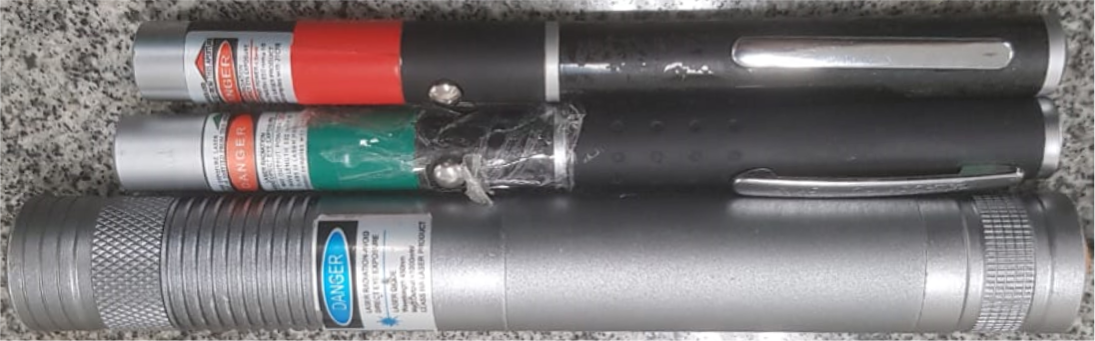}}
	\caption{Apparatus consisting of (a) an aquarium where a gradient of temperature was produced and (b) lasers of different frequencies/wavelenghts.}
	\label{fig:aparatotermico}
	\end{center}
\end{figure}

\subsection{Temperature Gradient}

Our apparatus consisted of the following materials:

\begin{description}
    \item [Plexiglass aquarium] $25\,\mathrm{cm}$ high and $40\,\mathrm{cm}$ long, filled with water;
    \item [Lasers of 3 different colors] red, green and blue, so that the wavelength of incident light could be varied;
    \item [Scale panel] used to set and measure the angle of incidence of light;
    \item [Resistors] to transfer heat to the water from the top of the aquarium;
    \item [Ice] to cool down water at the bottom;
    \item [Thermometers] to measure temperatures of water inside the aquarium at different heights;
    \item[Wooden stands] to hold the thermometers at chosen heights.
   \end{description}

We tried to create a model based on Fourier's law:

\begin{equation}
    \frac{dQ}{dt}=cA\frac{dT}{dy}
    \label{eq:FourierLaw}
\end{equation}
with a linear relation between height and temperature, i.e., with $dT/dy$ constant. 
For if temperature varies linearly with height, then height will be a linear function of temperature as well. 
Likewise, the index of refraction is known to vary linearly with temperature.
Therefore, height is expected to be a linear function of the index of refraction and vice-versa.

We performed an auxiliary experiment aiming to measure different values of $n$ in different layers of the gradient. 
However, we noticed that the values of $n$ for different temperatures were very similar, and therefore, the temperature gradient that we were able to produce could not provide conditions good enough to observe the light following a curved path. 
Still, we could observe that the light rays were bent after a while, but that was actually due to the wear and tear of the wooden stands that were holding the thermometers.
They were releasing sawdust into the water and this ended up forming a small gradient of concentrations, the real cause of the curvature. 
Thus, we realized that we would need to carry out our experiment in another way, and we decided to make a concentration gradient like what also occurs for this phenomenon in nature.

\begin{figure}[H]
    \centering
    \includegraphics[%
    width=0.3\columnwidth]{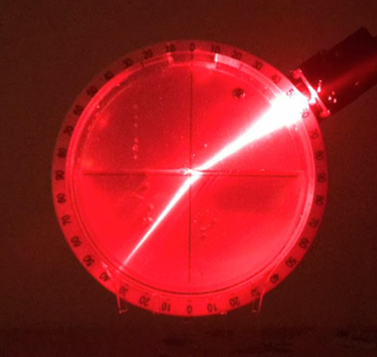}
    \caption{Auxiliary experiment performed to measure values of the refraction index for different layers of a temperature gradient.}
    \label{fig:auxiliartermico}
\end{figure}

\subsection{Gradient of Sugar Solution at Different Concentrations}

Our apparatus was almost the same previously mentioned. Besides the plexiglass aquarium, lasers, and scale to measure angles of incidence, we also used:

\begin{description}
    \item[51 Solutions of Sugar] graduated every $2\%$ at a time, with the standard solution taken as $100 \%$ of concentration corresponding to $1.5\,\mathrm{kg/}l$;
    \item [Burette] used to slowly and carefully insert each graduation of sugar solution;
    \item [Support] to fix the burette above the aquarium.
\end{description}

The solutions forming the concentration gradient are solutions of different percentages of sugar diluted or saturated in water, totaling 51 different solutions. 
The layers formed by different concentrations were slowly deposited, one by one, using the burette – a care we took to maximize the chances of obtain linearity of the gradient with height. 
The solutions were deposited from the most to the least concentrated one and the experimental gradient was built with 51 layers. 
Our theory, which considers a continuous gradient, also describes our case as the layers thickness are sufficiently small compared to the aquarium's dimensions.

\begin{figure}[H]
	\begin{center}
	\subfigure{\includegraphics[width=0.30\linewidth]{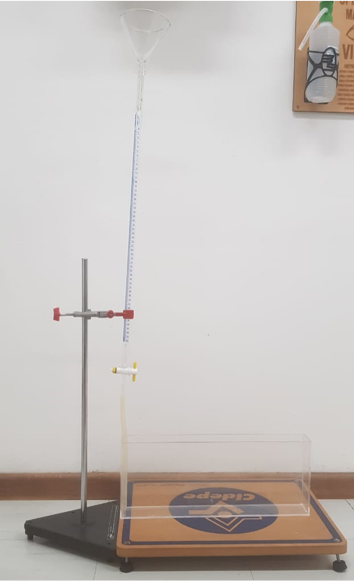}}
	\subfigure{\includegraphics[width=0.50\linewidth]{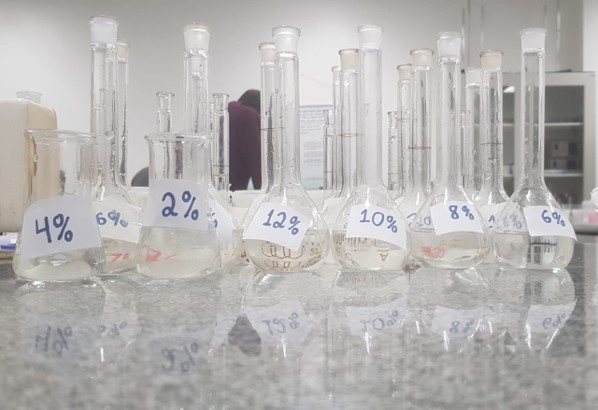}}
	\caption{Apparatus for (a) the concentration gradient (a), and (b) graduated sugar sulutions.}
	\label{fig:aparatoconcentracao}
	\end{center}
\end{figure}

By repeating the auxiliary experiment, we obtained a variation significantly bigger and now quantifiable in the values of $n$, which led us to carry on with this experiment. Theoretical aspects of the concentration gradient were described in Section~\ref{sec:fundamentos} and will be considered for our experimental analysis.

\begin{figure}[H]
	\begin{center}
	\subfigure{\includegraphics[width=0.30\linewidth]{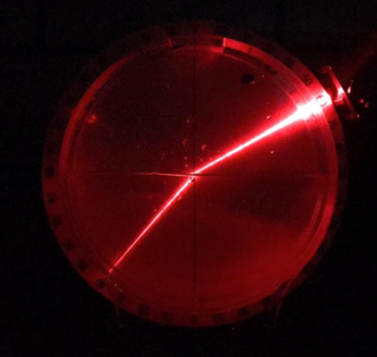}}
	\subfigure{\includegraphics[width=0.50\linewidth]{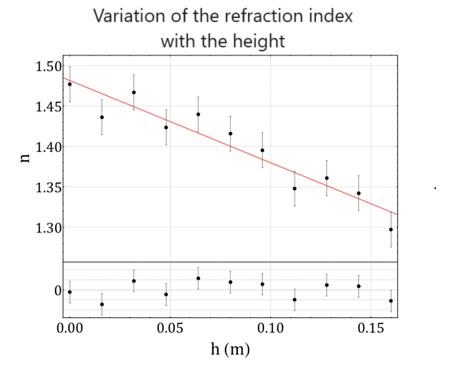}}
	\caption{Auxiliary experiment to measure values of the refraction index of different layers, and respective results.}
	\label{auxiliarconcentracao}
	\end{center}
\end{figure}

\section{Experimental Analysis}
\label{sec:analise}

In this section, we are going to present our fit to the expected path of light in relation to the points obtained with the aid of the \textit{Tracker} application. 
The main source of errors come from dissipation of light scattered by grains of sugar and also refraction when light passes through or reflects on the aquarium's walls. 
The distribution of errors and overall quality of this fit can be assessed by noticing that the residues do not show any particular trend. 
Although an increase in the errors is expected when light passes through lower layers, and 
therefore, the ray becomes thicker because there is an increasing number of particles 
scattering light, even considering only the uncertainties due to the size of each pixel, we 
notice that the residuals seem to be randomly scattered around zero and, therefore, we may state 
that random points were picked from the highlighted region, compensating accordingly. 
As a consequence, this will not affect any qualitative aspect of our analysis. 

\begin{figure}[H]
    \centering
    \includegraphics[%
    width=0.70\columnwidth]{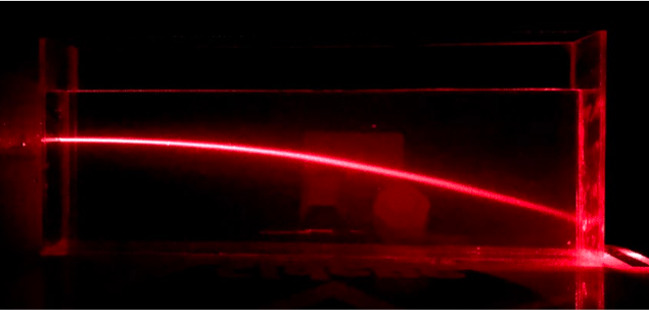}
    \caption{Example of image used to take points for the experimental analysis.}
    \label{fig:raiodeluz}
\end{figure}

\begin{figure}[H]
    \centering
    \includegraphics[%
    width=0.70\columnwidth]{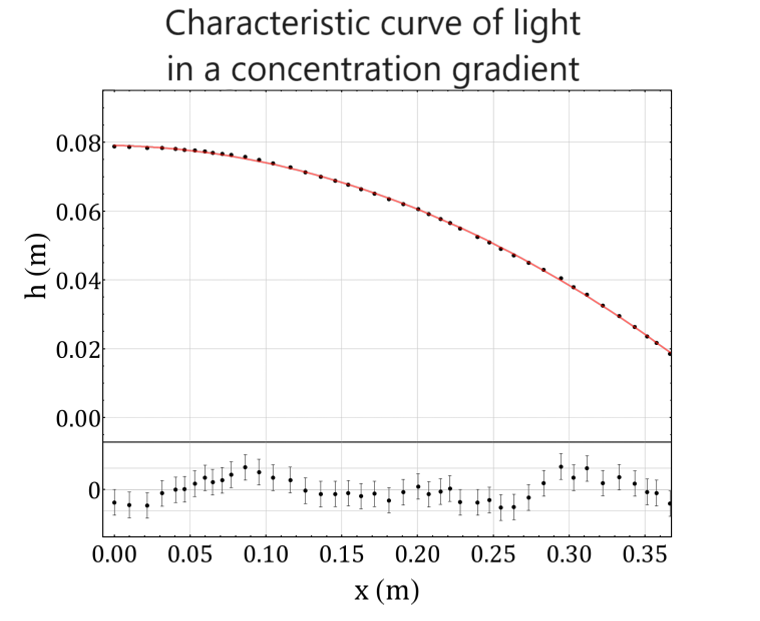}
    \caption{Experimental fit for the path followed by light in the concentration gradient and experimental points obtained from Figure~\ref{fig:raiodeluz}. The residues are presented under the graph.}
    \label{fig:graficocurva}
\end{figure}

\subsection{Reconstruction of Images}

Once we know the mathematical predictions for the path followed by each light ray, 
we can also write the corresponding equations for their tangent lines.
This allows us to identify the apparent positions from which the light rays would 
emerge under normal circumstances. 
Therefore, we can set conditions according to which we would observe the formation 
of an inverted image, which characterizes the occurrence of a Fata Morgana mirage.

\begin{figure}[H]
    \centering
    \includegraphics[%
    width=0.3\columnwidth]{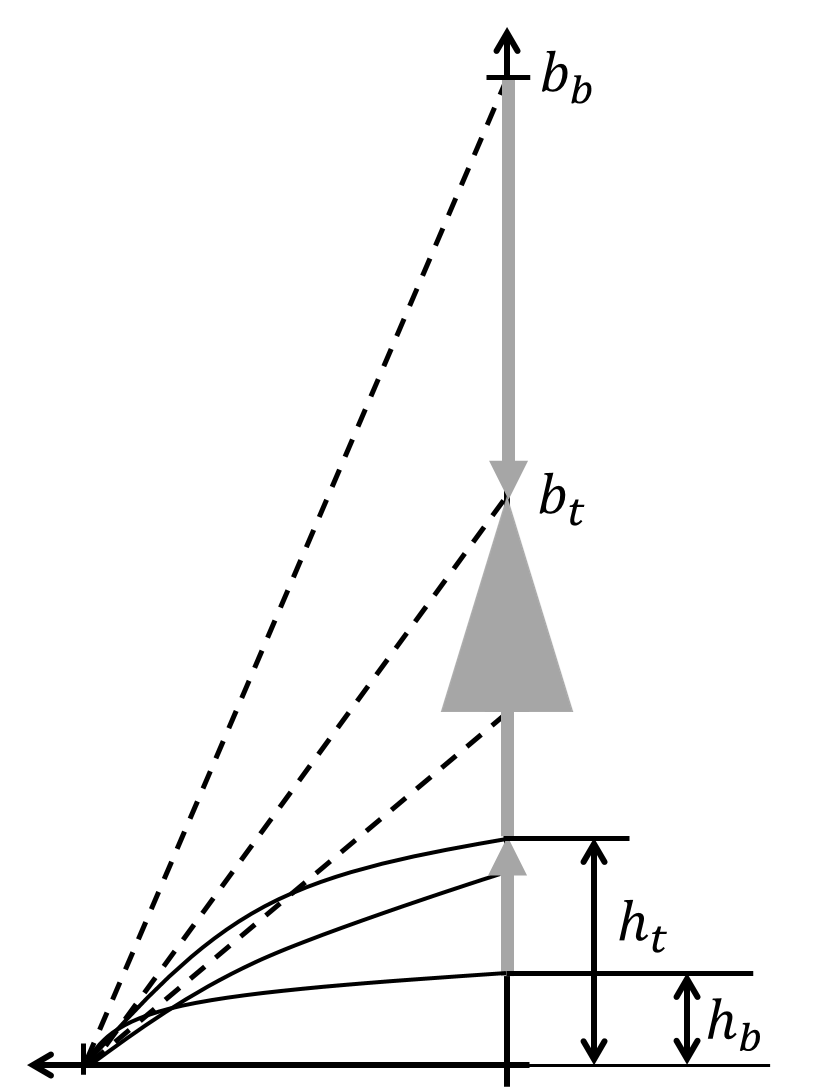}
    \caption{Scheme representing the formation of images in a Fata Morgana mirage. Notice that an inverted image occurs for $b_b > b_t$.}
    \label{fig:formacaoimagem}
\end{figure}

As we can notice from Figure~\ref{fig:formacaoimagem}, $y(x)$ will depend on $h_t$ and 
on a certain $\alpha_0$ for light coming from $y=h_t$. 
Let $x_0$ be the point receiving that incoming light. 
An observer at $x_0$ will perceive light coming from $y = h_t$, for instance, as it 
was actually travelling along a straight line directly from $y=b_t$. 
This straight line is tangent to the path defined by $y(x)$ at $x=x_0$  and its equation is:

\begin{equation}
    y = \left( y(x_0) - \frac{dy}{dx} (x=x_0)x_0 \right) 
    + \frac{dy}{dx} (x=x_0) x
    \label{linhaTangente}
\end{equation}
where $b = \left (y(x_0) - \frac{dy}{dx} (x=x_0)x_0 \right)$ is the linear coefficient 
(it will be $b_t$, if we consider light coming from $h_t$, for instance), and 
$a = \frac{dy}{dx} (x=x_0)$ is the angular coefficient, i.e., the inclination of 
this tangent line with respect to the horizontal.

In the same way, a function $y(x)$ representing the path of light coming from $y=h_b$ 
will depend on $h_b$ and a certain $\alpha_0$ and an equation like \eqref{linhaTangente} can be 
written for its tangent line, therefore defining the point from which light will seem to emerge for the observer at $x=x_0$. 

In order to study the formation of images and check if the experiment would really form a Fata 
Morgana, we placed a picture of Marie Curie on one side of the aquarium and positioned a camera 
on the opposite side to capture its image. 
We then tried to use the trajectories of the light rays obtained according to our model 
to construct a deformed image from the original one and then compare it with the image 
actually captured.

\begin{figure}[H]
	\begin{center}
	\subfigure{\includegraphics[width=0.20\linewidth]{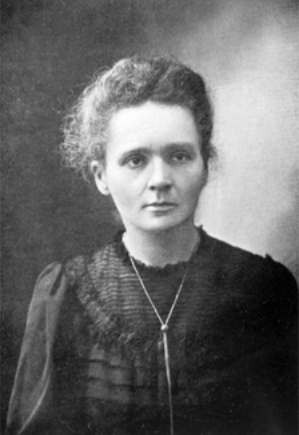}}
	\subfigure{\includegraphics[width=0.15\linewidth]{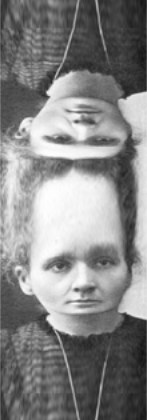}}
	\subfigure{\includegraphics[width=0.15\linewidth]{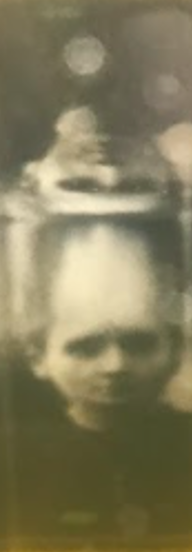}}
	\caption{(a) Original picture, (b) reconstructed image, and (c) captured image.}
	\label{fig:mariecurie}
	\end{center}
\end{figure}

In Figure~\ref{fig:mariecurie}, we see the original picture at left and, at right, the image that was 
captured through the aquarium. In the center,  we see the image reconstructed with the employment of our model.


It is possible to evaluate the similarity between reconstructed and captured images by highlighting some points of reference and comparing their positions on both images. In this sense, we found good agreement between our experimental observations and the corresponding predictions of our model.

\begin{figure}[H]
    \centering
    \includegraphics[%
    width=0.55\columnwidth]{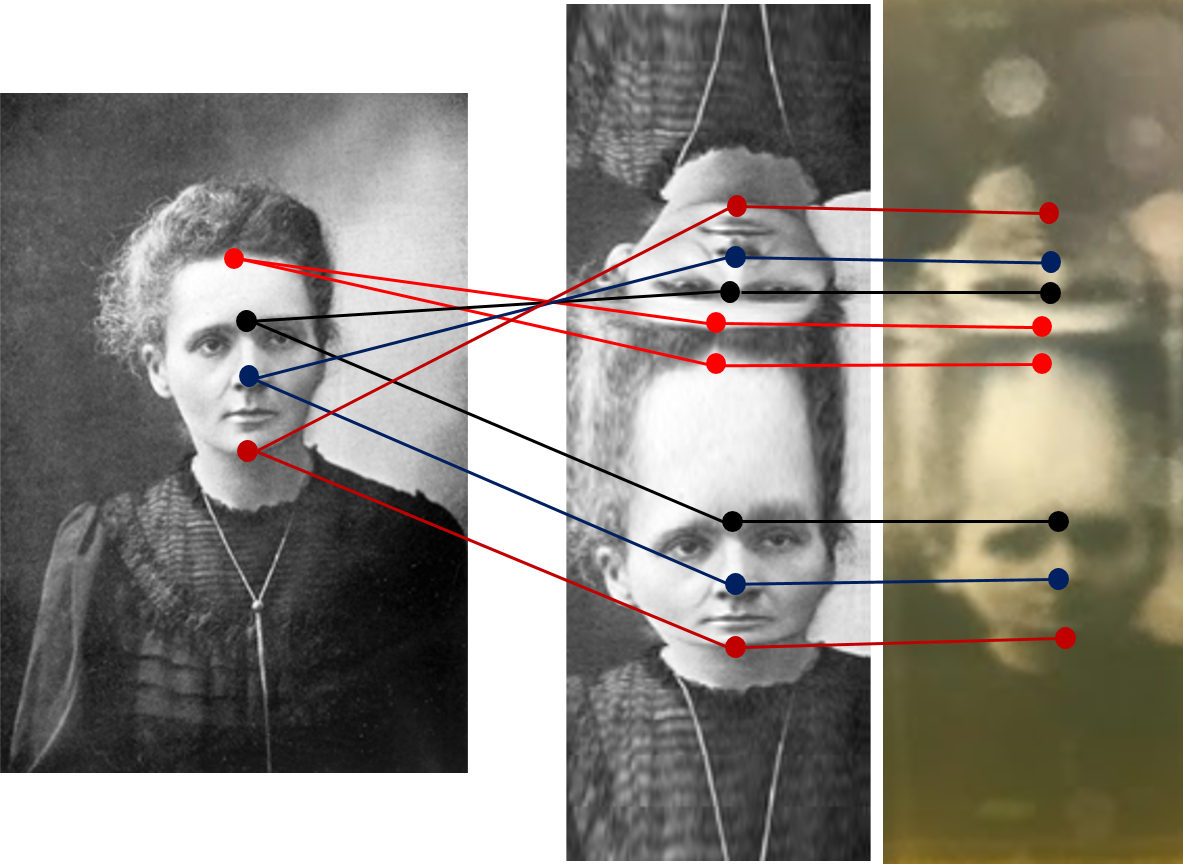}
    \caption{Comparison between points of reference of the original picture in reconstructed and captured images.}
    \label{fig:comparacaoimagem}
\end{figure}

\section{Conclusions}
\label{sec:conclusao} 

In this work, it was demonstrated with experiments that, within the scale of our 
apparatus, a concentration gradient has better practical effects for bending 
light rays than a thermal gradient. This is not necessarily expected as a general 
rule. For instance, when the Fata Morgana mirage is observed in nature, light will 
pass through atmospheric air with a combination of both, gradient and temperature 
gradients can occur, therefore constituting a more complex system for which a simplified analysis as the one presented here would not apply directly.
Despite of that, we defined an equation that can predict the path followed by light in a 
refraction index gradient, 
noting that such path depends on the first refracted angle of the light ray, the 
index of refraction of the 
first layer of incidence, the rate of change of the index of refraction with 
height and, also, the height at 
which incident light enters.
This part of our work involved approaching and working with 
elementary concepts of differential calculus, which is not a subject 
commonly taught in Brazilian high schools, except when private 
schools from a 
very restricted and privileged group offer specific training aiming  
international exams, as the Advanced Placement (AP), or scientific 
Olympiads, as the IYPT. On the other hand, pre-calculus courses, as 
they are popularly known, are not something so rare to be found 
as part of the curricula of final years in schools throughout the 
world. 
In the case presented here, almost nothing exceeding common 
curriculum was really taught in expository classes, but actually 
resulted from research performed by students followed by group 
discussions and guidance.
Also, no more than low cost materials were used and, in many cases, important 
parts of our apparatus were handmade by ourselves, an example being the 
rectangular aquarium. Those materials later became part of the school's 
patrimony and are now available to be used to reproduce this and other 
experiments.
In the final part of this project, it was possible to 
theoretically and experimentally establish image 
reconstruction conditions and 
to perform such reconstruction for a specific image in lab conditions, 
which allowed us to compare a captured distorted image with another 
one, that was constructed following our theoretical prediction. This 
final step constituted an important visual appeal linked to the 
understanding of the phenomenon beyond formulas and graphs, 
reaching a meaningful connection with initial expectations and 
also improving the scientific communicability associated with the 
general idea of this project. In this sense, we believe the 
developments presented here could be useful as an example of an active 
learning procedure allowing, not only an early introduction to some 
relatively advanced mathematical concepts and methods in scientific 
research, but mainly, a meaningful tool for the engagement of young 
students, the first author of this paper being one of them, who is now 
attending Amherst College and pursuing degrees in both Physics and 
Mathematics.

\appendix

\section{\label{subsec:A} Influence of the Wavelength}
Using the Sellmeier equation, we obtained values for our lasers' wavelengths $\lambda$:

\begin{equation*}
    n^2(\lambda) = 1+\sum_i \frac{Bi\lambda^2}{\lambda^2 - D}
\end{equation*}
where $B$ and $D$ are the Sellmeier coefficients.

\begin{table}[h]
\centering
\caption{Refraction indexes $n$ for water at room temperature}
\vspace{0.5cm}
\begin{tabular}{l|c|c|c}

Color & Wavelength ($nm$)  & Absolute refraction index & $n_0 / k$ \\
\hline
Red & $650 \pm 10$  & $1.3388$ & $1.367 \pm 0.009$\\
Green & $532 \pm 10$  &  $1.3337$ & $1.334 \pm 0.004$ \\
Blue & $405 \pm 10$  & $1.3310$ & $1.306 \pm 0.003$

\end{tabular}
\end{table}

We only obtained variations in the third decimal place, and due to our experimental uncertainty in $n_0 / k$, being $k$ the inclination of the line that relates the index of refraction and the height, also occurring in the third decimal place, we therefore have that differences in $n$ for the different colors used can be ignored.


\begin{thebibliography}{99}
\bibitem{Young2}A.T. Young, \emph{An Introduction to Mirages}, (1999)
\bibitem{Young1}A.T. Young, E. Frappa, \emph{Mirages at Lake Geneva: the Fata Morgana}, Applied Optics \textbf{56},  19 (2017)
\bibitem{Exner} J.M. Pernter,  F.M. Exner, \emph{Meteorologische Optik}, (Wein und Leipzig, W. Braumuller, 1922),  2nd ed., p. 163–188
\bibitem{IYPT}International Young Physicists' Tournament, \emph{Problems for the 33rd IYPT 2020} (2019)
\end{thebibliography}
\end{document}